\documentclass[twocolumn,prb,showpacs,preprintnumbers,amsmath,amssymb,superscriptaddress,reprint]{revtex4-1}[11pt]
\usepackage[pdftex]{graphicx}
\usepackage{amsmath}
\usepackage{amsfonts}
\usepackage{amssymb}
\usepackage{subfigure}
\usepackage{color}
\usepackage{float}
\usepackage{lipsum}
\usepackage{color}
\begin{document}
\title{Resonant spin transfer torque nano-oscillators}
\author{Abhishek Sharma}
\author{Ashwin. A. Tulapurkar}
\author{Bhaskaran Muralidharan}
\affiliation{Department of Electrical Engineering, Indian Institute of Technology Bombay, Powai, Mumbai-400076, India}
\date{\today}
\medskip
\widetext
\begin{abstract}
	Spin transfer torque nano-oscillators are potential candidates for replacing the traditional inductor based voltage controlled oscillators in modern communication devices. Typical oscillator designs are based on trilayer magnetic tunnel junctions which are disadvantaged by low power outputs and poor conversion efficiencies. In this letter, we theoretically propose to use resonant spin filtering in pentalayer magnetic tunnel junctions as a possible route to alleviate these issues and present device designs geared toward a high microwave output power and an efficient conversion of the d.c. input power. We attribute these robust qualities to the resulting non-trivial spin current profiles and the ultra high tunnel magnetoresistance, both arising from resonant spin filtering. The device designs are based on the nonequilibrium Green's function spin transport formalism self-consistently coupled with the stochastic Landau-Lifshitz-Gilbert-Slonczewski's equation and the Poisson's equation.  We demonstrate that the proposed structures facilitate oscillator designs featuring a large enhancement in microwave power of around $775\%$ and an efficiency enhancement of over $1300\%$ in comparison with typical trilayer designs. We also rationalize the optimum operating regions via an analysis of the dynamic and static device resistances. This work sets stage for pentalyer spin transfer torque nano-oscillator device designs that extenuate most of the issues faced by the typical trilayer designs.  
\end{abstract}
\maketitle
\indent Spin transfer torque nano-oscillators (STNOs) are a class of non-linear nanoscale oscillators which have attracted a lot of interest from the physics as well as the applications perspective. The interest from the physics perspective stems from the need to advance the understanding of magnetization dynamics in non-linear systems\cite{Tiberkevich2007,Kim2006,Kim2008,Slavin2009,Kim2012}. From the applications perspective, these devices find suitability in the modern communication electronics \cite{Katine2008,Wolf2006,Choi2014}. STNOs have better in-built features over traditionally used voltage control oscillators (VCOs), such as smaller size, lower cost and easier integrability to silicon technology. In order to technologically replace VCOs, STNOs should be able to deliver high microwave power outputs and must possess higher conversion efficiencies with a good quality factor. There have been consistent efforts \cite{Covington2004,Deac2008,Nano-oscillators2012} to improve the performance of STNOs based on typical trilayer magnetic tunnel junctions (MTJ). Various improvements proposed are centered around modifying the magnetic properties of the ferromagnet (FM). However, they have not been able to deliver microwave power outputs in excess of $0.3\mu$W \cite{Nano-oscillators2012}. In this work we propose pentalayer device designs that make use of resonant spin filtering, termed as resonant tunneling magnetic tunnel junction (RTMTJ) structures, to circumvent these issues faced by typical trilayer based STNO designs. 
\begin{figure}[htb!]
	\subfigure[]{\includegraphics[width=1.95in]{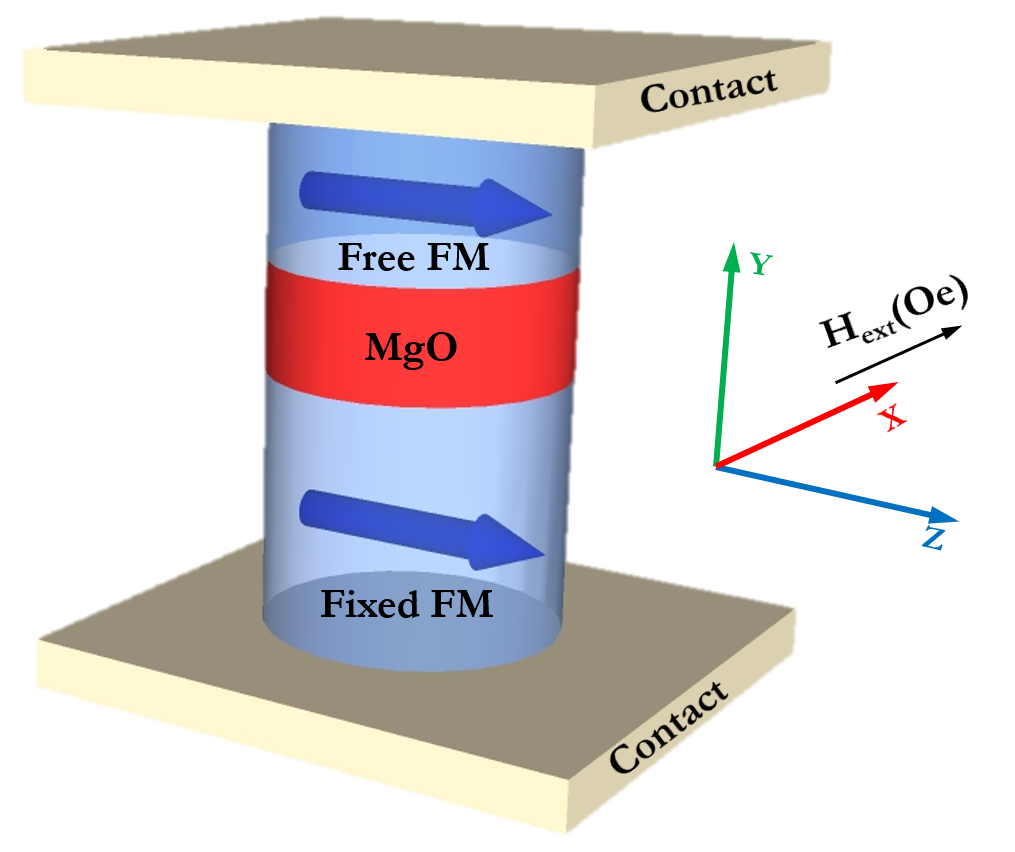}		
	}\subfigure[]{\includegraphics[width=1.6in]{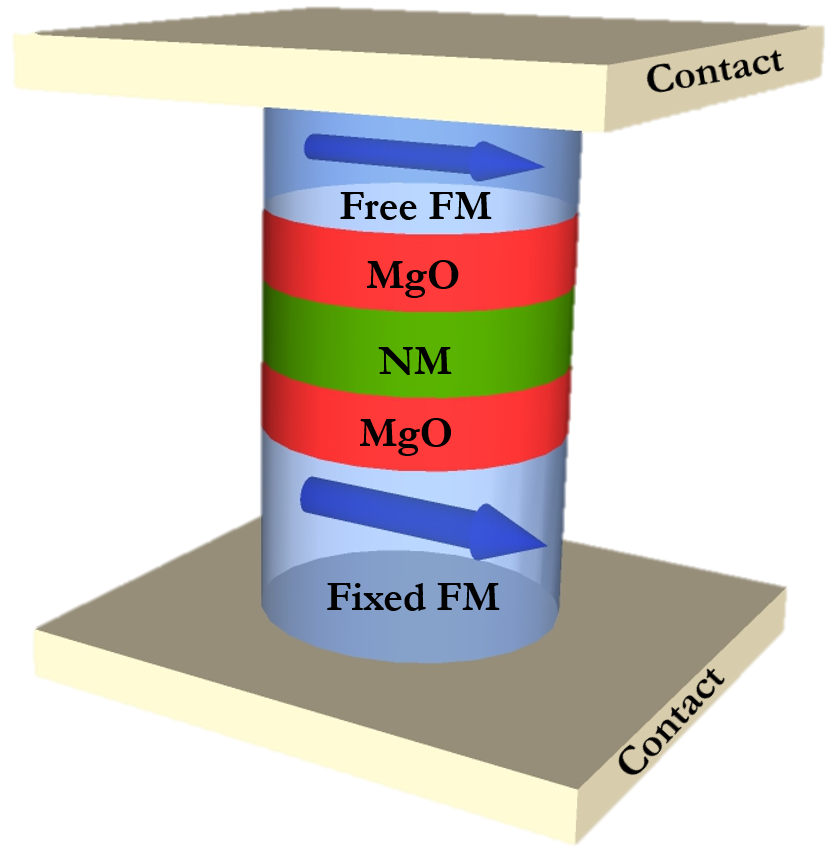}}
	\subfigure[]{\includegraphics[width=2.7in]{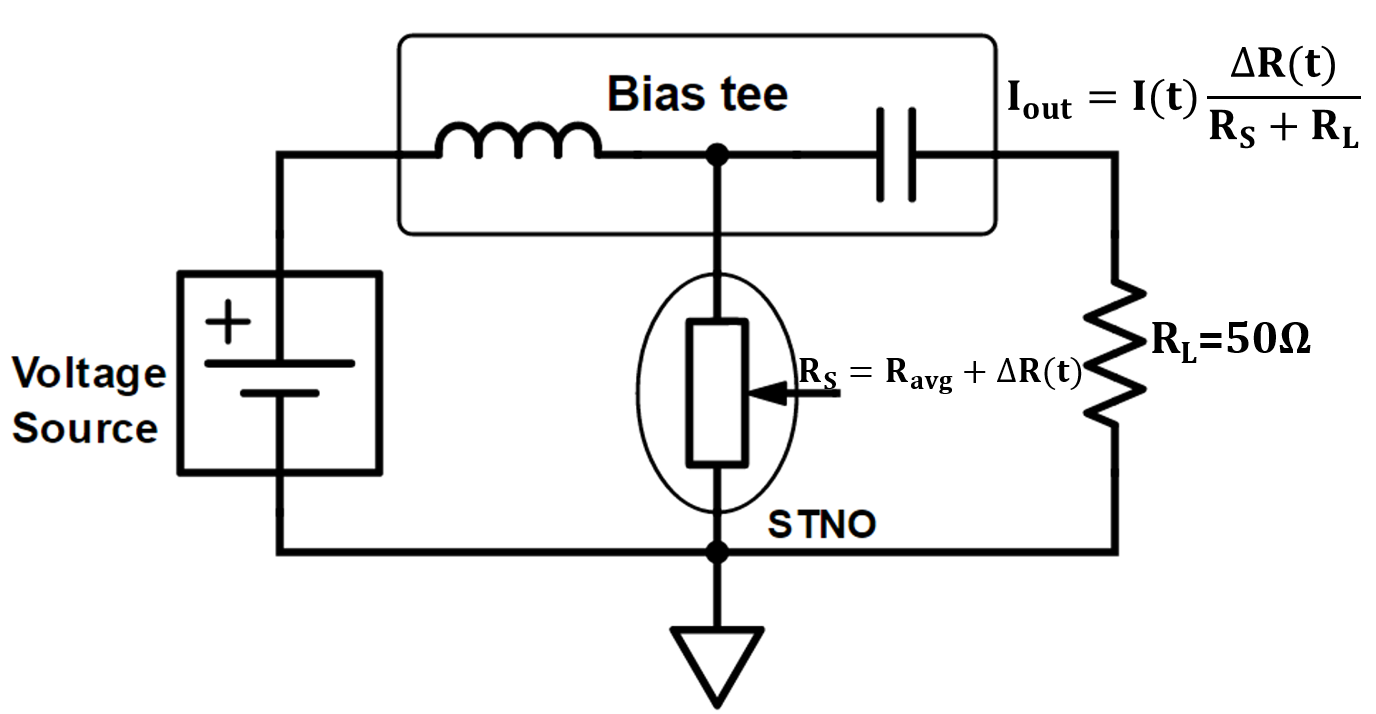}}
	\caption{Oscillator design schematics. (a) Typical trilayer device with an insulating MgO layer between the free and the fixed ferromagnetic layers, (b) RTMTJ-based device comprises a MgO-Normal metal(NM)-MgO heterostructure between the free and the fixed ferromagnetic layers. An external field magnetic field ($H_{ext}$) is applied  along  $\hat{x}$-direction.(c) Circuit diagram of an STNO biased by a voltage source with the microwave power delivered to a load resistor $R_L$.}
	\label{device_design}
\end{figure}
We demonstrate that owing to the novel spin-filtering physics in the proposed structures \cite{Chatterji2014a,Sharma2016}, the resulting non-trivial spin current profiles and the high tunnel magneto resistance (TMR) translate to an ultra improvement in the STNO performance.\\
\indent Spin transfer torque \cite{Berger1996,Slonczewski1996} involves the transfer of spin angular momentum from spin-polarized charge carriers to the magnetization of the ferromagnetic layer. Spin torque can either enhance the magnetic damping inherent in magnetic systems or can compensate for the damping processes, based on the state of the ferromagnet and the direction of the spin-polarized current. When the spin torque magnitude is large enough to compensate magnetic damping, an instability in the magnetization dynamics results. In MTJs, the magnetic state of the free ferromagnet can be switched either parallel or anti-parallel with respect to the pinned FM layer (see Fig.~\ref{device_design}(a)), due to spin torque under a sufficient voltage bias. The state of the free ferromagnet can be toggled back to the initial state by applying a static magnetic field. This results in self-sustained oscillations of the magnetization in the free ferromagnetic layer. The nature of the self sustained oscillations is governed by the magnetization dynamics incited by the spin current profile. These self sustained oscillations in the magnetization translate to high-frequency electrical signals due to the magneto resistance (MR) effect. The microwave power output thus translated, is directly associated with the electrical readout (i.e., the MR) and the ratio $I/I_C$\cite{Slavin2009}, where $I$ is the bias current and $I_C$ is the critical current required for magnetization switching.  \\
\indent One may anticipate an increase in the power output by ramping the ratio $I/I_C$, which can be achieved at a higher voltage bias. However, a higher bias in turn reduces the MR of the device as evidenced in experiments as well as our simulation results (see Fig.~\ref{RT_TMR_V}(b)), ultimately resulting in a reduction in the microwave output power.  Therefore, high microwave power outputs through STNOs can be achieved by designing a device that combines high MR and low switching bias. Various studies have focussed on lowering $I_C$ by tailoring the magnetic properties of the ferromagnetic layer while preserving the higher MR. They have estimated that the maximum power delivered to a matched load is around $1\mu$W, while the maximum achieved power in experiments is still around $0.3\mu$W\cite{Nano-oscillators2012}. In this work, we thus propose to harvest the higher MR and the lower switching bias emerging from resonant spin filtering physics \cite{Chatterji2014a,Sharma2016,RTMTJ_2015} to increase the microwave power and the conversion efficiency of STNOs.\\
\begin{figure}[htb!]
	\subfigure[]{\includegraphics[width=1.75in]{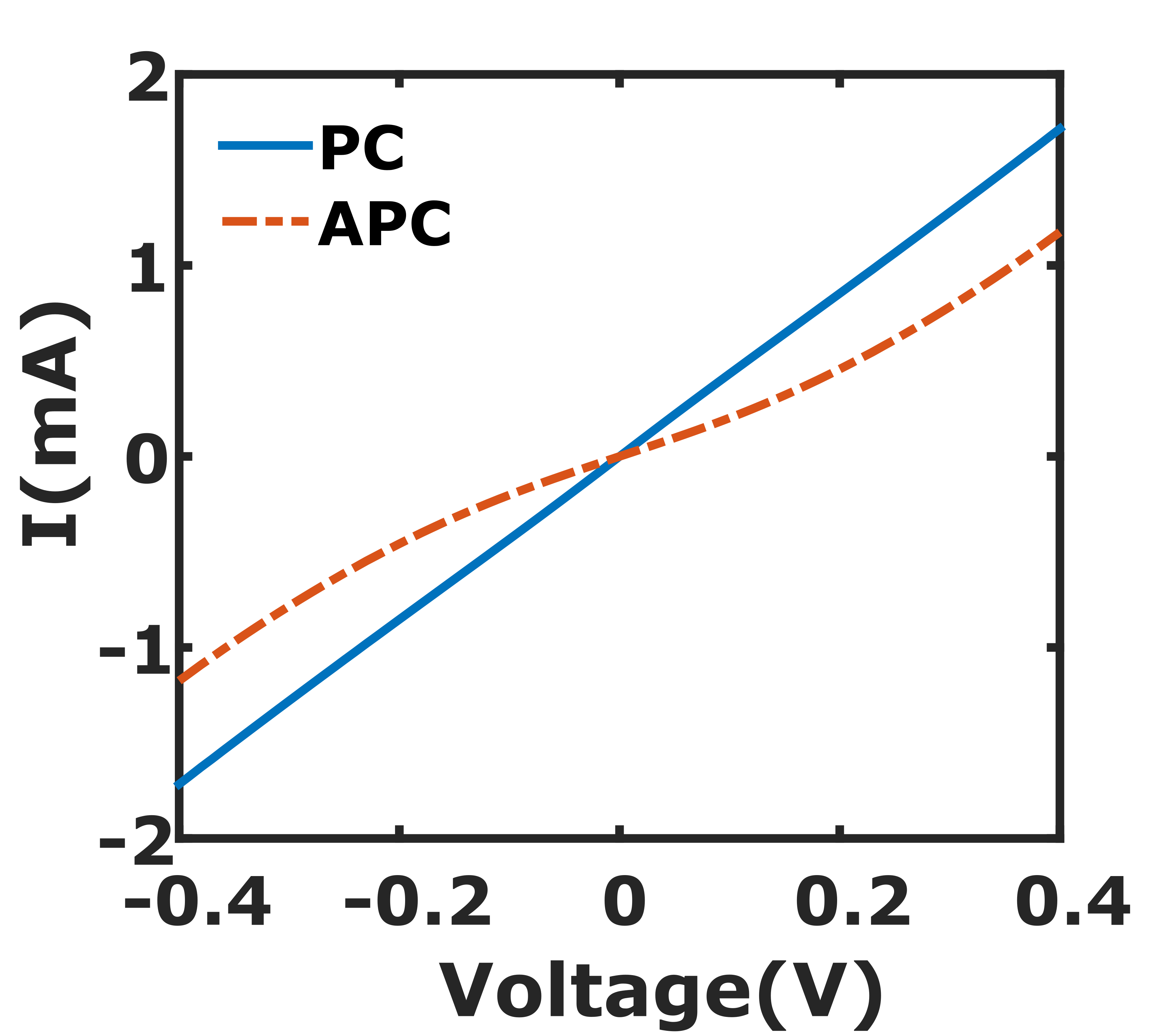}		
	}\subfigure[]{\includegraphics[width=1.85in]{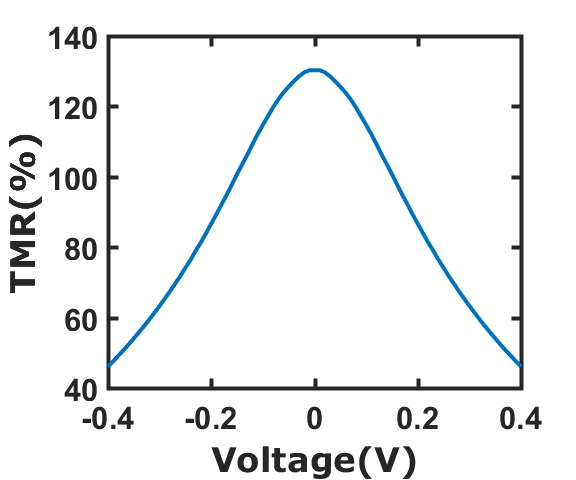}}
	\subfigure[]{\includegraphics[width=1.8in]{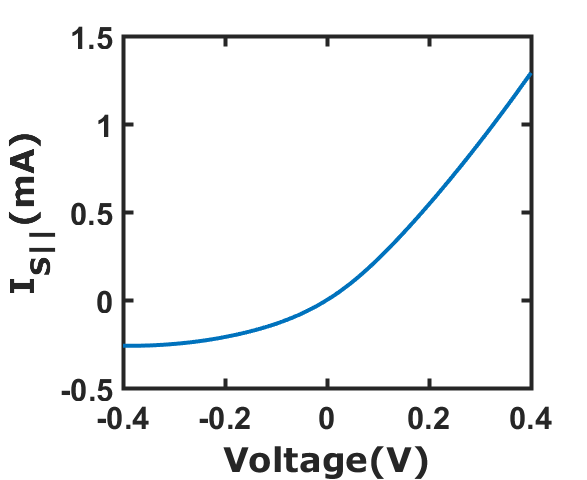}	
	}\subfigure[]{\includegraphics[width=1.8in]{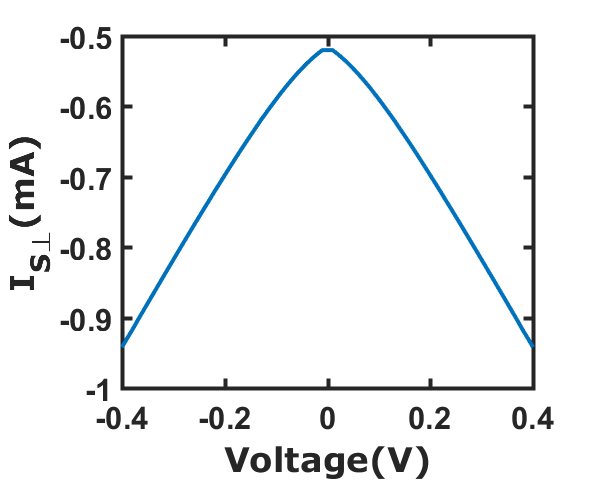}}
	\caption{Trilayer device characteristics: (a) The current variation with bias voltage in the parallel configuration(PC) and the anti-parallel configuration(APC), (b) TMR variation with voltage, (c) The $I_{S\parallel}$ (Slonczewski term) and, (d)  The $I_{S\perp}$ (field-like term) variations with voltage in the PC and the APC conflagrations. Note that the two are coincident on each other in this case.}
	\label{TMR_V}
\end{figure}
\indent Device schematics for both the trilayer and the pentalayer structures are depicted in Fig.~\ref{device_design}(a), and Fig.~\ref{device_design}(b) respectively. The equivalent circuit is schematized in Fig.~\ref{device_design}(c). These designs have been simulated by employing the non-equilibrium Green's function (NEGF)\cite{datta1} spin transport formalism coupled with the Poisson's equation and the Landau-Lifshitz-Gilbert-Slonczewski (LLGS)\cite{Slonczewski1996} equation, as described in our earlier works \cite{Chatterji2014a,Sharma2016} (also see supplementary material). In this work, we have also taken into account the thermal noise in the form of magnetic field fluctuations $\vec{h_r}$ in the LLGS equation with the following statistical properties \cite{Garcia-Palacios1998}
\begin{equation}
\langle h_{fl,i}(t)\rangle = 0, \langle h_{fl,i}h_{fl,j}(s)\rangle = 2D\delta_{i j}\delta(t-s)
\end{equation}
where i and j are Cartesian indices, and $\langle \rangle$ represents the ensemble average. The strength of the fluctuation $D$ is given by  
\begin{equation}
D=\frac{\alpha}{1+\alpha^2}\frac{k_BT}{\gamma\mu_0M_SV}
\end{equation}\\ 
where, $\alpha$ is the Gilbert damping parameter, $\gamma$ is the gyro-magnetic ratio of the electron, $\mu_0$ is the free space permeability constant, $k_B$ is the Boltzmann constant, $T$ is the temperature of the magnetic layer, $M_S$ and $V$ are the saturation magnetization and the volume of the free layer respectively.\\
\indent In our simulations, we use CoFeB as the ferromagnet with its Fermi energy, $E_f = 2.25$eV and exchange splitting $\Delta = 2.15$ eV. The effective mass of MgO is $m_{OX} = 0.18m_e$ and of the normal metal, $m_{NM} = 0.36m_e$, with $m_e$ being the free electron mass. The barrier height of the CoFeB-MgO interface is $U_B = 0.76$ eV above the Fermi energy \cite{deepanjan,kubota}. \\
\indent In the results that follow, the parameters chosen for the magnetization dynamics are $\alpha$ = 0.01, the saturation magnetization $M_S=1200$ emu/cc, $\gamma$ = 17.6 MHz/Oe, with the anisotropy field $H_k=75$Oe along $\hat{z}$-axis,  which have been extracted form Z. Zeng et. al., \cite{Nano-oscillators2012} after removing the zero bias field and the demagnetization field of $H_d=1500$Oe\cite{Nano-oscillators2012} along $\hat{y}$-axis. The cross-sectional area of all the devices considered is 70 $\times$ 160 nm\textsuperscript{2} with thickness of the free ferromagnetic layer taken to be 1.6 nm. All the simulations have been done at room temperature. The RTMTJ structure shown in Fig.~\ref{device_design}(b) may be realized either by an appropriate non-magnetic metal sandwiched between MgO barrier\cite{Korenivski2008} or via a heterostructure of MgO and a stoichiometrically substituted MgO $(\mbox{Mg}_\text{{x}}\mbox{Zn}_\text{{1-x}}\mbox{O})$, whose bandgap and workfunction may be tuned \cite{Li2014}. \\
\indent We show in Fig.~\ref{TMR_V}(a), the current-voltage (I-V) characteristics of a trilayer device in the parallel configuration (PC) and in the anti-parallel configuration (APC). The charge current is smaller in magnitude in the APC in comparison to the PC due to spin dependent tunneling in a trilayer device. The tunnel magneto resistance (TMR) is defined as $TMR=(R_{AP}-R_P)/(R_P)$, where $R_P$ and $R_{AP}$ are the resistances in the parallel and in the anti-parallel configurations, respectively. The TMR variation with the voltage for a trilayer device is shown in the Fig.~\ref{TMR_V}(b). Figure ~\ref{TMR_V}(c) shows the variation of the Slonczewski term \cite{butler} ($I_{S\parallel}$) of the spin current  (see supplementary information--Theoretical formulation) with bias voltage. The Slonczewski term can either enhance the damping in the magnetization dynamics or can compensate for the damping processes in the magnetic system, regulated by the direction of current. It can be seen from the Fig.~\ref{TMR_V}(d) that the field like term\cite{butler} ($I_{S\perp}$) of the spin current is non-zero at zero bias. This zero-bias component is a dissipation-less spin current and represents the exchange coupling between the ferromagnets due to the tunnel barrier\cite{Slonczewski1996}. This exchange coupling can be either ferromagnetic or anti-ferromagnetic in nature determined by the relative positioning of the conduction bands in the ferromagnets and the insulator. The exchange coupling is of anti-ferromagnetic nature in MgO based trilayer devices. The field like term serves as the effective magnetic field in the magnetization dynamics. In case of a trilayer device, it can be seen from the Fig.~\ref{TMR_V}(c), that $I_{S\parallel}$ has similar bias characteristics in both the PC as well as the APC. Similarly, it can be seen from the Fig.~\ref{TMR_V}(d), that $I_{S\perp}$ is identical for the PC and the APC in the trilayer case. \\ 
\indent The RTMTJ device has an ultra high TMR as shown in the Fig.~\ref{RT_TMR_V}(b) which can be tuned via appropriate positioning of the transmission peaks with respect to the Fermi level and the ferromagnetic exchange splitting $\Delta$ \cite{Sharma2016}. The resonant conduction in the PC and the off-resonant conduction in the APC (Fig.~\ref{RT_TMR_V}(a)) are responsible for the ultra high TMR \cite{Sharma2016}.  The larger Slonczewski term $I_{S\parallel}$ in the RTMTJ device as shown in the Fig.~\ref{RT_TMR_V}(c), can be attributed to the resonant conduction and enhanced spin filtering \cite{Sharma2016}. We show in Fig.~\ref{RT_TMR_V}(d) the variation of  $I_{S\perp}$ (field like term) with voltage. Here, it is interesting to note that the zero bias the exchange coupling is ferromagnetic in nature for the RTMTJ structure and an applied bias tries to change this exchange coupling to anti-ferromagnetic. Thus, at some applied bias it is possible to decimate the exchange coupling in the RTMTJ structure.\\
\begin{figure}[htb!]
	\subfigure[]{\includegraphics[width=1.75in]{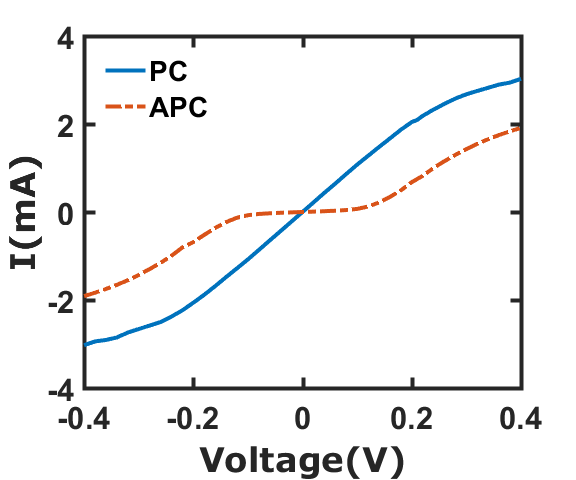}		
	}\subfigure[]{\includegraphics[width=1.8in]{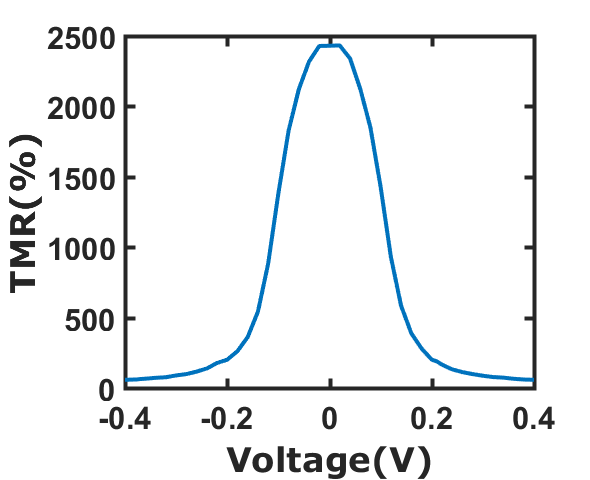}}
	\subfigure[]{\includegraphics[width=1.75in]{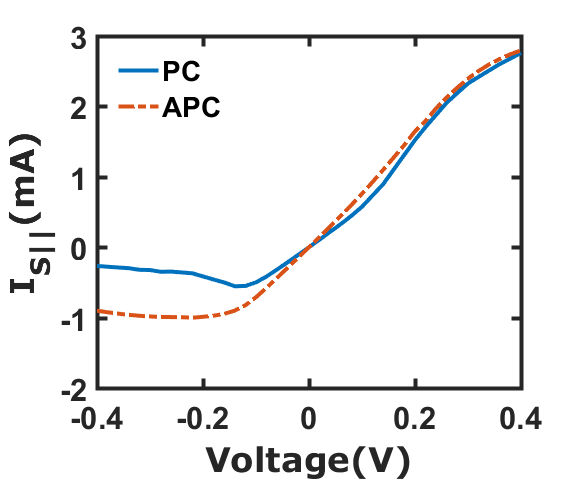}	
	}\subfigure[]{\includegraphics[width=1.75in]{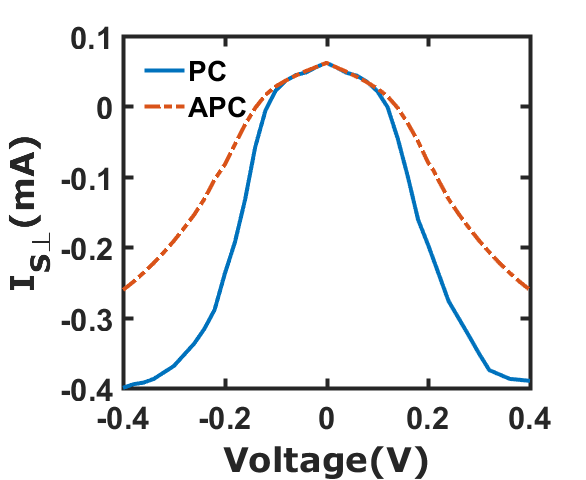}}
	\caption{Pentalayer device (RTMTJ) characteristics: (a) Current variation with bias voltage, (b)  TMR variation with bias voltage, (c)  $I_{S\parallel}$ (Slonczewski term) variation with bias voltage, (d)  $I_{S\perp}$ (field-like term) variation with voltage in the parallel configuration(PC) and the anti-parallel configuration(APC).}
	\label{RT_TMR_V}
\end{figure}
\indent In the case of STNOs, the non-linearity parameters can be varied over a wide range by changing the orientation and magnitude of the applied magnetic field \cite{Slavin2009}. When the orientation and magnitude of the external field in the plane of magnetization (see Fig.~\ref{device_design}(a)) is varied, we noticed that the external magnetic field perpendicular to the easy axis leads to high microwave power outputs and narrow line widths, consistent with an earlier theoretical work \cite{Slavin2009}.\\ 
\indent Based on the circuit diagram shown in Fig.~\ref{device_design}(c), we model the STNO as a source of time varying resistance connected with a $50\Omega$ load resistance. The power delivered to the load resistance constitutes the useful microwave power that can be extracted from the STNO and is given by:\\
\begin{equation}
P_{ac}=R_L \mbox{Var}\left(\frac{R_S(t)I_S(t)}{R_S(t)+R_L}\right)
\label{acPower}
\end{equation}
where $R_L=50\Omega$, $R_S(t)=V/I_S(t)$, is the source resistance and `Var' is the variance of the time dependent term. We show in Fig.~\ref{MTJ_tune}(a) the microwave power as a function of voltage in the trilayer device when an in-plane field is applied perpendicular to the easy axis of the free ferromagnetic (FM) layer. It is noted that the microwave power increases with applied bias due to the large spin current (Fig.~\ref{TMR_V}(c)). This results in large amplitude peak-to-peak magnetization dynamics translating to a large microwave power output. However, with further increase in bias voltage, the microwave power starts to fall off due to the reduction in the TMR at higher voltages (Fig.~\ref{TMR_V}(b)). We show in Fig.~\ref{MTJ_tune}(b) the variation of central frequency ($fc$) of microwave oscillations with bias for a trilayer device. 
\begin{figure}[htb!]
	\subfigure[]{\includegraphics[width=1.75in]{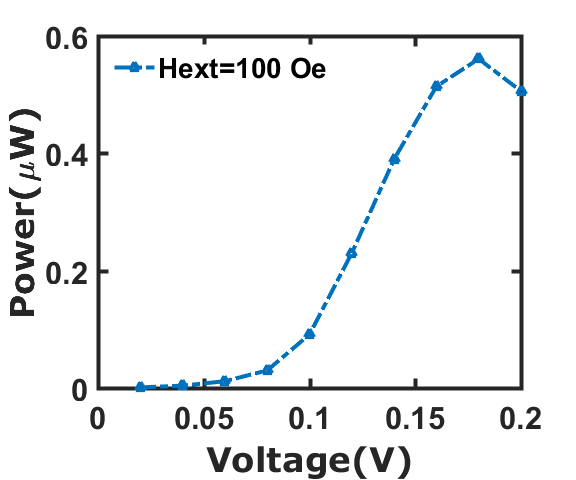}		
	}\subfigure[]{\includegraphics[width=1.8in]{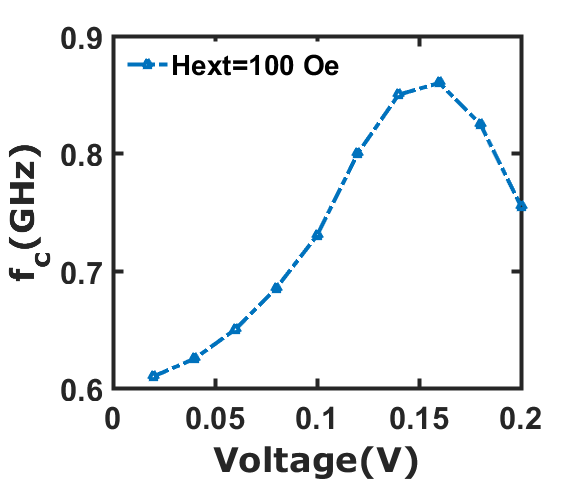}}
	\subfigure[]{\includegraphics[width=1.75in]{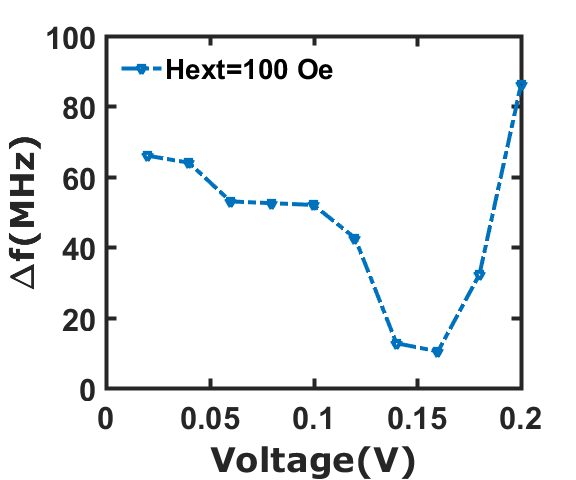}	
	}\subfigure[]{\includegraphics[width=1.75in]{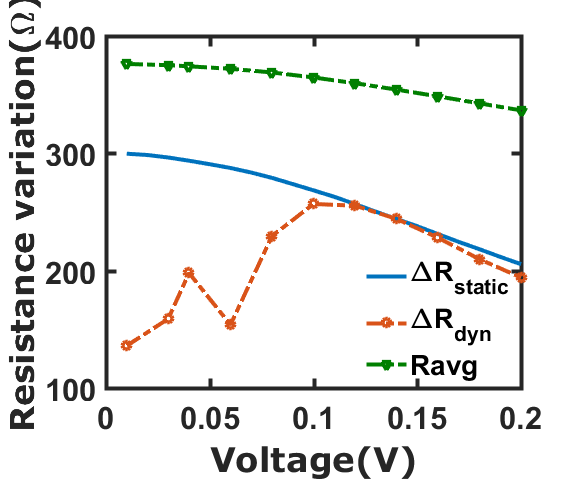}}
	\caption{Voltage-induced precession of a trilayer-MTJ device: (a) Voltage dependence of microwave power delivered to the $50\Omega$ load, (b) peak frequency, (c) FWHM(full width half maxima) $\Delta$f and (d) Resistance variation ($\Delta R_{static}$, $\Delta R_{dynamic}$, $R_{avg}$) as a function of the bias voltage.}
	\label{MTJ_tune}
\end{figure}
\begin{figure}[htb!]
	\subfigure[]{\includegraphics[width=1.75in]{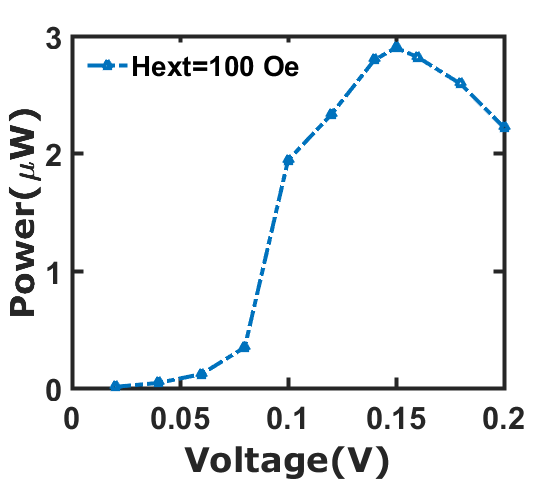}		
	}\subfigure[]{\includegraphics[width=1.8in]{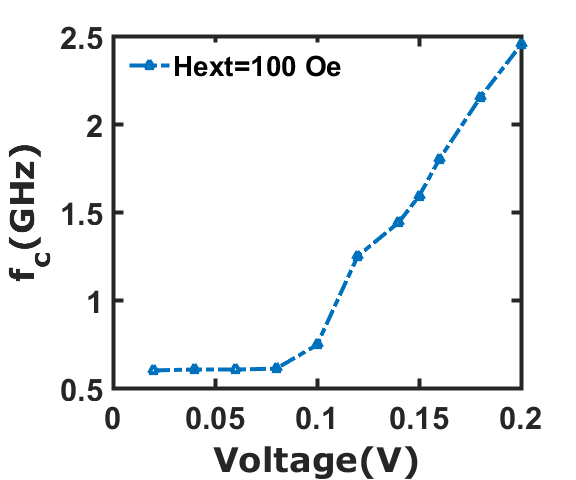}}
	\subfigure[]{\includegraphics[width=1.75in]{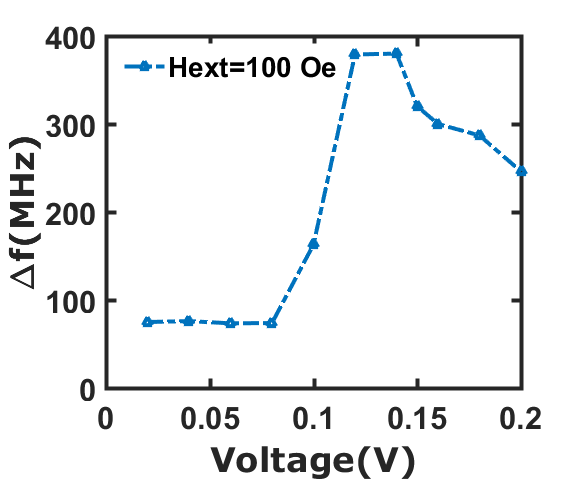}	
	}\subfigure[]{\includegraphics[width=1.75in]{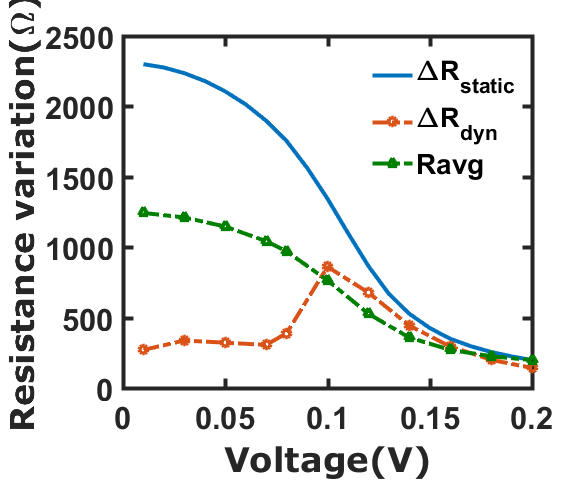}}
	\caption{Voltage-induced precession of the RTMTJ device: (a) Voltage dependence of microwave power delivered to the $50\Omega$ load, (b) peak frequency and (c) FWHM(full width half maxima) $\Delta$f (d) Resistance variation ($\Delta R_{static}$, $\Delta R_{dynamic}$, $R_{avg}$) as a function of the applied bias voltage.}
	\label{RTMTJ_tune}
\end{figure}
These trends in the power output and $fc$ can be understood by analyzing how the dynamic resistance ($\Delta R_{dynamic}$), the static resistance ($\Delta R_{static}$) and the average resistance ($Ravg$) vary with voltage as shown in the Fig.~\ref{MTJ_tune}(d). The dynamic resistance is the maximum change in the resistance of the device as the power oscillates, i.e.,  $\Delta R_{dynamic}=V/I_{min}-V/I_{max}$ \cite{Houssameddine2007}. The static resistance is the change in the resistance due to the MR effect, i.e., $\Delta R_{static}=R_{AP}(V)-R_P(V)$\cite{Houssameddine2007}. With increase in bias, $\Delta R_{dynamic}$ approaches $\Delta R_{static}$ as can be seen in Fig.~\ref{MTJ_tune}(d). This signifies large peak-to-peak magnetization dynamics and out of plane oscillations (OP) of the free ferromagnetic layer\cite{Houssameddine2007} (see supplementary material Fig.~3). The point of peak microwave power (Fig.~\ref{MTJ_tune}(a)) is shifted by a small amount from the point where $\Delta R_{dynamic}$ approaches $\Delta R_{static}$, due to the loading effect of $R_L$. As the microwave power delivered to the load increases when the load and source have the same resistances, any reduction in $Ravg$ increases the microwave power. However, with further increase in the bias, $\Delta R_{dynamic}$ starts to deviate from $\Delta R_{static}$, as seen in Fig.~\ref{MTJ_tune},  resulting in a reduction of microwave output power. It can be seen that the central frequency (Fig.~\ref{MTJ_tune}(b)) also peaks around the same voltage point where the $\Delta R_{dynamic}$ approaches the $\Delta R_{static}$. The frequency of oscillations ($f_C$) is determined by the demagnetization field with $f_C \propto m_y$, where $m_y$ is the out-of-plane component of the magnetization unit vector. Therefore, $f_C$ increases as $\Delta R_{dynamic}$ approaches $\Delta R_{static}$ associated with a higher component of the out-of-plane magnetization (see supplementary material Fig.~3). As the bias is further increased, the reduction in the $\Delta R_{dynamic}$ is more in comparison to $\Delta R_{static}$ resulting in a smaller out-of-plane magnetization component which further causes the central frequency to fall at a higher voltage. Figure~\ref{MTJ_tune}(c) shows the line width of the microwave signal as a function of bias voltage for the trilayer device. It is observed that the line width falls to $11$MHz at $V=0.16$, delivering $0.51\mu$W power to the $50\Omega$ load at a central frequency of $860$MHz.\\
\indent The microwave power for the RTMTJ based oscillator has a similar trend as that of a trilayer based oscillator, as can be seen in Fig.~\ref{RTMTJ_tune}(a). The frequency of oscillations is higher in the RTMTJ based oscillator due to the larger spin currents (see Fig.~\ref{RTMTJ_tune}(b)) in comparison with the trilayer device. The line width is also larger in this device as shown in Fig.~\ref{RTMTJ_tune}(c), but has nearly the same quality factor as that of the trilayer device. The central frequency for the RTMTJ device increases monotonically with voltage as the $\Delta R_{dynamic}$ steadily approaches the $\Delta R_{static}$ (Fig.~\ref{RTMTJ_tune}(d)) making the device more suitable for the high frequency applications.\\
\begin{figure}[htb!]
	\subfigure[]{\includegraphics[width=1.75in]{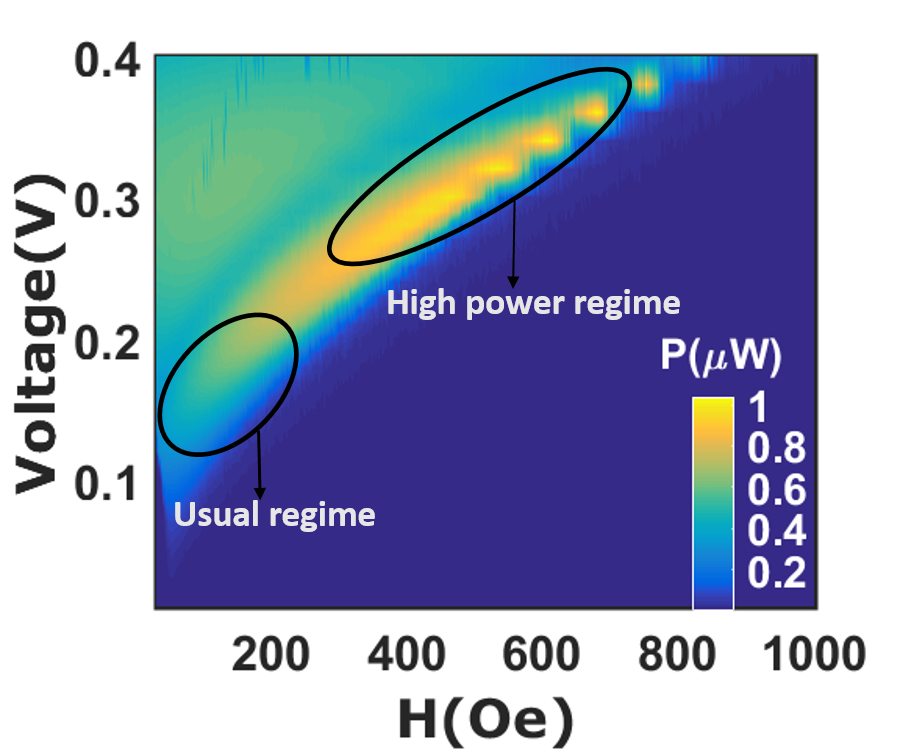}		
	}\subfigure[]{\includegraphics[width=1.75in]{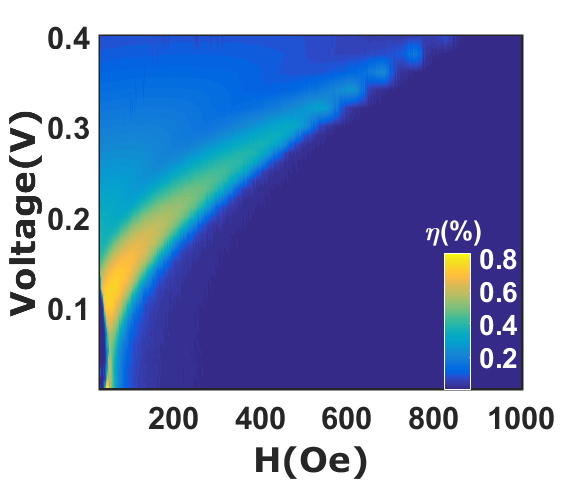}}
	\caption{Voltage-field diagram of the trilayer-MTJ device for (a) Microwave power delivered to $50\Omega$ load, (b) conversion efficiency $\eta(\%)$.}
	\label{MTJ_Pout}
\end{figure}
\begin{figure}[htb!]
	\subfigure[]{\includegraphics[width=1.75in]{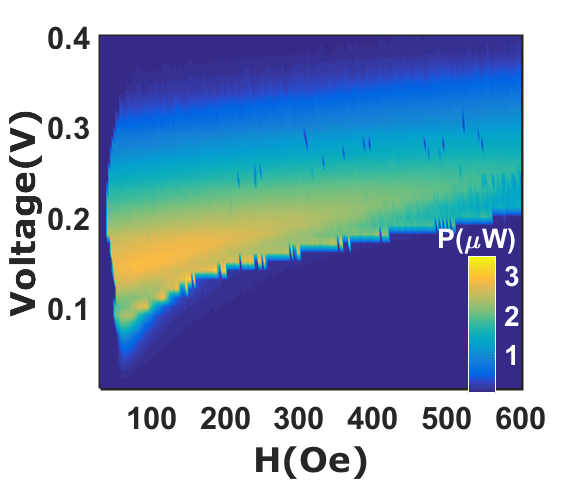}		
	}\subfigure[]{\includegraphics[width=1.75in]{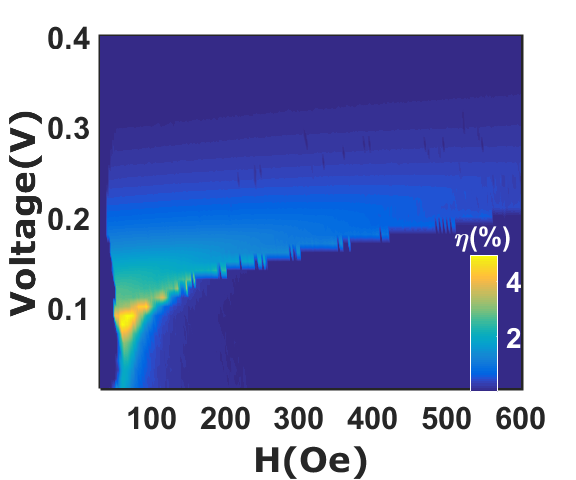}}
	\caption{Voltage-field diagram of the RTMTJ device for (a) Microwave power delivered to $50\Omega$ load, (b) conversion efficiency $\eta(\%)$.}
	\label{RTMTJ_Pout}
\end{figure}
Further, it can be seen from Fig.~\ref{MTJ_Pout}(a) that the microwave power delivered to the load by the trilayer device has two operating regimes marked as the `Usual regime' and the `High power regime'. The maximum power delivered to  the load in the `Usual regime' is around $0.5\mu$W. In the `High Power regime' the  microwave output power is nearly $1\mu$W under the bias of $V=0.34$ and the external field of $H_{ext}=605$Oe. The conversion efficiency, i.e., $\eta=P_{a.c.}/P_{input}$) of the trilayer based oscillator at the maximum microwave output power point is $0.23\%$. The high power outputs in this regime can be associated with comparable dynamic and static resistances ($\Delta R_{dynamic}=132\Omega$, $\Delta R_{static}=132.1\Omega$) in conjunction with a small average resistance ($R_{avg}=299\Omega$). Due to the high spin current in the `high power regime', the frequency of oscillations is higher in comparison to the `usual regime'. At maximum power point in `high power regime', the frequency of oscillations is $f_C=2.45GHz$. It can seen from the Fig.~\ref{MTJ_Pout}(b) that the efficiency of a trilayer based oscillator is high in the `Usual regime' in comparison to the `high power regime' due to the small input voltage bias.\\
\indent Figure~\ref{RTMTJ_Pout}(a) shows the microwave power delivered to the $50\Omega$ load by an RTMTJ based oscillator. The RTMTJ based oscillator has two major features namely the high output power (see Fig.~\ref{RTMTJ_Pout}(a)) and the ultra high conversion efficiency (see Fig.~\ref{RTMTJ_Pout}(b)) in comparison to the trilayer based oscillator device (see Fig.~\ref{MTJ_Pout}(a) \& (b)). The maximum power delivered to the $50\Omega$ load is $3.5\mu$W which occurs at $V=0.13$V and an external field $H_{ext}=157$Oe. The efficiency of the RTMTJ device at maximum power is $3.23\%$. Hence, the RTMTJ based oscillator delivers $250\%$ higher power and is $1300\%$ more efficient in comparison to the trilayer based oscillator operating in `high power regime'. Further, the RTMTJ based device oscillator delivers $775\%$ more power in comparison to the trilayer device operating in the `usual regime'.\\
\indent We have thus proposed and explored designs of STNOs based on resonant tunneling to harvest two of its special features i.e.,  the ultrahigh TMR and the capability to exhibit large spin currents at small bias voltages. We have demonstrated that the resonant spin filtering of the RTMTJ makes the structure most suitable candidate for the next generation STNOs from the device perspective. We have estimated that the STNOs based on the RTMTJ device deliver $775\%$ higher microwave power with $1300\%$ better efficiency in comparison to the trilayer-MTJ based oscillator. We believe that this demonstration of RTMTJ as an oscillator will open up new frontiers for  experimental considerations of pentalayer structures and theoretical investigations of spin feedback oscillators\cite{Kumar2016,Bhuktare2017} based on such structures. This can pave way for the next generation STNOs in modern communications\cite{Choi2014}.\\
{\it{Acknowledgements:}} The author Abhishek Sharma would like to acknowledge Smarika Kulshrestha for her insightful discussions. This work was in part supported by the IIT Bombay SEED grant and the Department of Science and Technology (DST), India, under the Science and Engineering Board grant no. SERB/F/3370/2013-2014.\\
\newpage
\section{Supplementary Information on ``Resonant spin transfer torque nano-oscillators''}
\section{Theoretical Formulation}
\begin{figure*}[htb!]
	\subfigure[]{\includegraphics[width=3in]{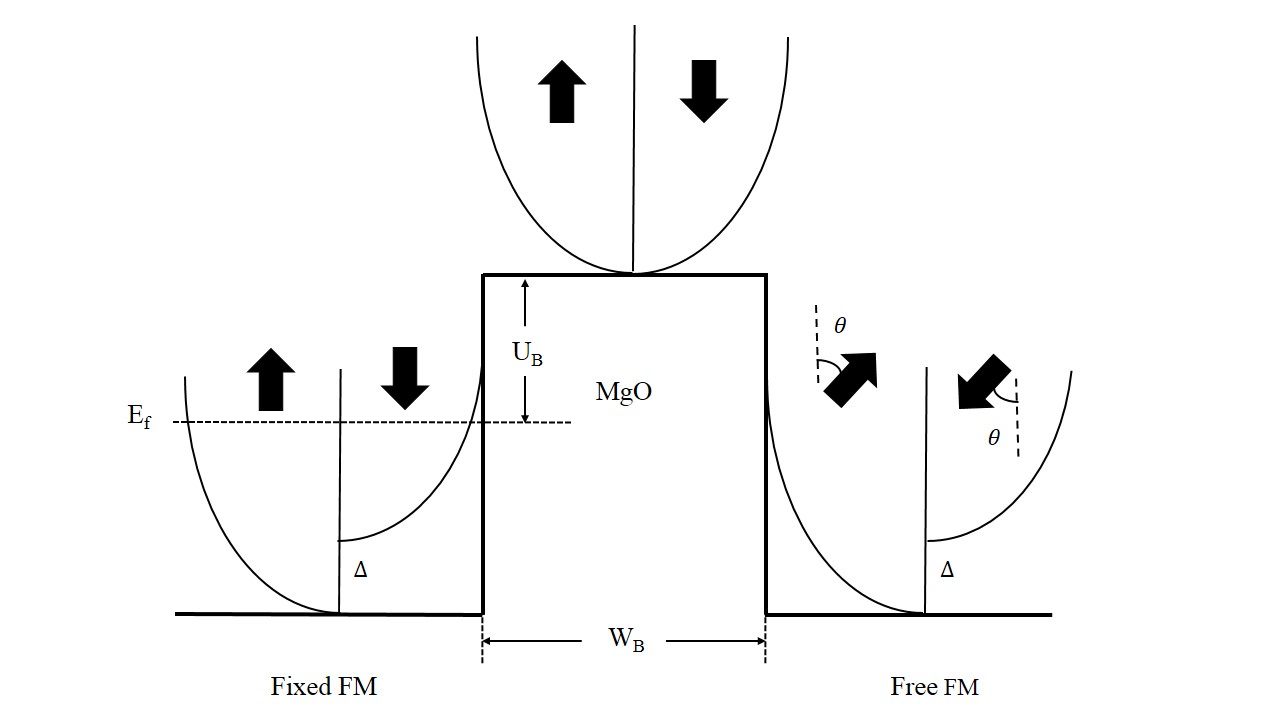}		
	}\subfigure[]{\includegraphics[width=3in]{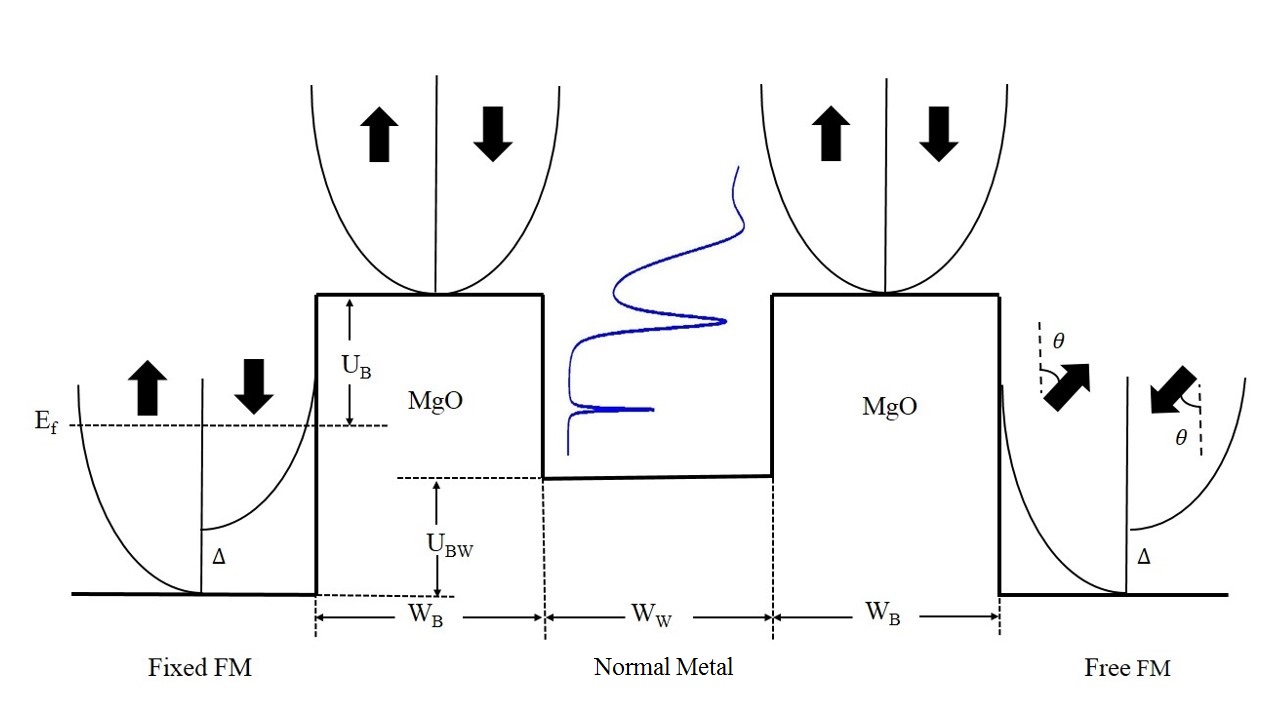}}
	\caption{Energy band schematic. (a) A trilayer MTJ device at equilibrium along $\hat{z}$ direction. The ferromagnetic contacts have an exchange energy of $\Delta$ with $E_f$ being Fermi energy, $U_B$, the barrier height in MgO above Fermi Energy. (b) An RTMTJ device at equilibrium along $\hat{z}$ directoin. Here, $U_{BW}$ is the difference between the bottom of the conduction band of the ferromagnet and the normal metal or semiconductor}
	\label{device_band}
\end{figure*}

\begin{figure*}[ht]
	\centering
	\includegraphics[width=\linewidth]{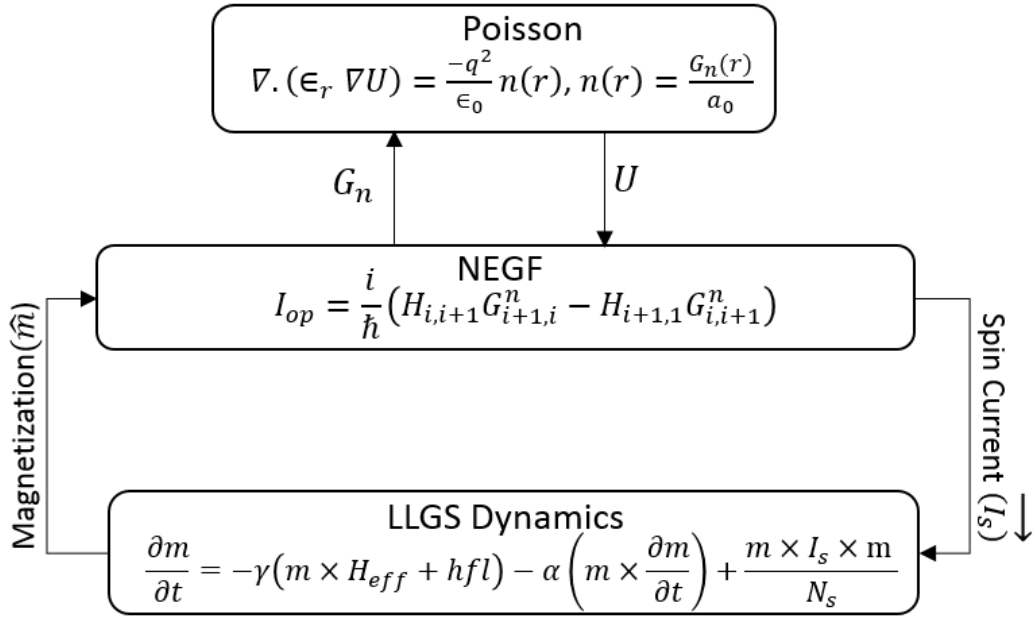}
	\caption{Simulation engine for nonequilibrium Green'€™s function spin transport formalism self-consistently coupled with the stochastic Landau-Lifshitz-Gilbert-Slonczewski's and the Poisson's equation}
	\label{simulation_engine}
\end{figure*}

We sketch the essential details of the non-equilibrium Green's function (NEGF) simulation procedure \cite{butler,yanik,deepanjan,lunds,akshay} that was used to analyze the nano-oscillator device designs, based on the device structures detailed in Fig.~\ref{device_band}. The trilayer MTJ has a layer of MgO between the magnets while the RTMTJ has a heterostructure of MgO-Normal metal-MgO sandwiched between the fixed and the free magnets leading to resonant peaks in the transmission spectrum. The magnetization of the fixed layer is along the $\hat{z}$-axis and that of the free layer changes with an applied bias and magnetic field.\\
\indent The NEGF spin transport formalism self-consistently coupled with the stochastic Landau-Lifshitz-Gilbert-Slonczewski's (LLGS) and the Poisson's equation within the effective mass framework is employed to calculate the charge and spin currents in the devices \cite{datta2,butler,lunds,akshay,yanik} as shown in Fig.~\ref{simulation_engine}. We start with the energy resolved spin dependent single particle Green's function matrix $[G(E)]$ evaluated from the device Hamiltonian matrix $[H]$ given by:
\begin{equation}
[G(E)] = [EI-H-\Sigma]^{-1}
\label{green_fun}
\end{equation}
\begin{eqnarray}
[\Sigma]=[\Sigma_T]+[\Sigma_B],
\label{Sigma}
\end{eqnarray}
where the device Hamiltonian matrix,  $[H]=[H_0]+[U]$, comprises the device tight-binding matrix, $[H_0]$ and the Coulomb charging matrix ,$[U]$, in real space, $[I]$ is the identity matrix with the dimensionality of the device Hamiltonian. The quantities $[\Sigma_T]$ and $[\Sigma_B]$ represent the self-energy matrices \cite{datta2} of the top and bottom magnetic layers evaluated within the tight-binding framework \cite{yanik,deepanjan}.  A typical matrix representation of any quantity $[A]$ defined above entails the use of the matrix element $A(z,z',k_x,k_x',k_y,k_y',E)$, indexed on the real space $z$ and the transverse mode space $k_x,k_y$. To account for the finite cross-section, we follow the uncoupled transverse mode approach, with each transverse mode indexed as $k_x,k_y$ evaluated by solving the sub-band eigenvalue problem \cite{lunds,salah2,deepanjan}. \\
\indent The charging matrix, $[U]$, is obtained via a self consistent calculation with the Poisson's equation along the transport direction $\hat{z}$ given by
\begin{eqnarray}
\frac{d}{dz}\left(\epsilon_r(z)\frac{d}{dz} U(z)\right)=\frac{-q^2}{\epsilon_0}n(z) \label{poisson}\\
n(z)=\frac{1}{A.a_0}\displaystyle\sum_{k_x,k_y}G^n(z;k_x,k_y),
\label{n_r}
\end{eqnarray}
with $G^n(z;k_x,k_y)=G^n(z,z,k_x,k_x,k_y,k_y)$, being a diagonal element of the energy resolved electron correlation matrix $[G^n(E)]$ given by
\begin{equation}
[G^n]= \int dE [G(E)] [\Sigma^{in}(E)] [G(E)]^{\dagger}	\label{G_n}
\end{equation}
\begin{equation}
[\Sigma^{in}(E)]=[\Gamma_T(E)]f_T(E)+[\Gamma_B(E)]f_B(E) \label{Sigma_in},
\end{equation}
Here, $[\Gamma_T(E)]=i\left ([\Sigma_T(E)]- [\Sigma_T(E)]^{\dagger} \right )$ and $[\Gamma_B(E)]=i\left( [\Sigma_B(E)]-[\Sigma_B(E)]^{\dagger} \right )$ are the spin dependent broadening matrices \cite{datta2} of the top and bottom contacts. The Fermi-Dirac distributions of the top and bottom contacts are given by $f_T(E)$ and $f_B(E)$ respectively. Here, $U(z)$ is the potential profile inside the device subject to the boundary conditions, $U_{FixedFM}=-qV/2$ and $U_{FreeFM}=qV/2$, with $V$ being the applied voltage, $A$ being the cross sectional area of the device, $a_0$ being the inter-atomic spacing in effective mass framework and $\hbar$ being the reduced Planck's constant. \\
\indent The summit of the calculation is the evaluation of charge currents following the self-consistent convergence of \eqref{poisson} and \eqref{n_r}. The matrix element of the charge current operator $\hat{I}_{op}$ representing the charge current between two lattice points $i$ and $i+1$ is given by \cite{datta1}
\begin{equation}
{I}_{op,i,i}=\frac{i}{\hbar}\left(H_{i,i+1}G^{n}_{i+1,i}-G^{n\dagger}_{i,i+1}H^{\dagger}_{i+1,i}\right) ,
\end{equation}
following which the charge current $I$ and spin current $I_S$ are given by  $I =q \int dE \text{ Real [Trace(}\hat{I}_{op}\text{)]}$, $I_{S\sigma} =q \int dE \text{ Real [Trace(}\sigma_S\cdot\hat{I}_{op}\text{)]}$ respectively where, the current operator $\hat{I}_{op}$ is a 2$\times$2 matrix in spin space, $H$ is the Hamiltonian matrix of the system and $q$ is the electronic charge.\\
\indent We have resolved spin current as $\vec{I_S}=I_{S,m}\hat{m}+I_{S,\parallel}\hat{M}+I_{S,\perp}\hat{M}\times\hat{m}$, the $I_{S\parallel}$ along  $\hat{M}$ is known as Slonczewski spin transfer torque term and the $I_{S\perp}$ along $\hat{M}\times\hat{m}$ is known as field like term. We use the Landau-Lifshitz-Gilbert-Slonczewski (LLGS) equation to calculate the magnetization dynamics of the free layer in the presence of an applied magnetic field and spin current. \cite{slon,brat}:
\begin{widetext}
\[
\left( 1+\alpha^{2}\right) \frac{\partial \hat{m}}{\partial t} = -\gamma \hat{m} \times (\vec{H}_{eff}+\vec{h_{fl}}) - \gamma \alpha \left( \hat{m} \times ( \hat{m} \times (\vec{H}_{eff}+\vec{h_{fl}}))\right)
- \frac{\gamma\hbar}{2qM_SV}[(\hat{m}\times(\hat{m}\times\vec{I_S}))-\alpha(\hat{m}\times\vec{I_S})]
\nonumber
\]
\end{widetext}
where $\hat{m}$ is the unit vector along the direction of magnetization of the free magnet, $\gamma$ is the gyromagnetic ratio of the electron, $\alpha$ is the Gilbert damping parameter, $\vec{H}_{eff} = \vec{H}_{app} + H_km_{z}\hat{z} - H_dm_{x}\hat{x}$ is the effective magnetic field with $\vec{H}_{app}$ being the applied external field, $H_k=\frac{2K_{u\parallel}}{M_S}$ being the anisotropy field, and $H_d=4\pi M_s-\frac{2K_{u\perp}}{M_S}$ being effective demagnetization field, $K_{u\parallel}$, $K_{u\perp}$ being in-plane and perpendicular uni-axial anisotropy constant respectively, $M_S$ is the saturation magnetization of free layer, with V being the volume of free ferromagnetic layer.In this work, we have also taken into account the thermal noise in the form of magnetic field fluctuations $\vec{h_r}$ in the LLGS equation with the following statistical properties \cite{Garcia-Palacios1998}
\begin{equation}
\langle h_{fl,i}(t)\rangle = 0, \langle h_{fl,i}h_{fl,j}(s)\rangle = 2D\delta_{i j}\delta(t-s)
\end{equation}
where i and j are Cartesian indices, and $\langle \rangle$ represents the ensemble average. The strength of the fluctuation $D$ is given by  
\begin{equation}
D=\frac{\alpha}{1+\alpha^2}\frac{k_BT}{\gamma\mu_0M_SV}
\end{equation}\\ 
where, $\mu_0$ is the free space permeability constant, $k_B$ is the Boltzmann constant, T is the temperature of the magnetic layer.\\
\section{Magnetization dynamics}
\begin{figure*}[htb!]
	\subfigure[]{\includegraphics[width=2in]{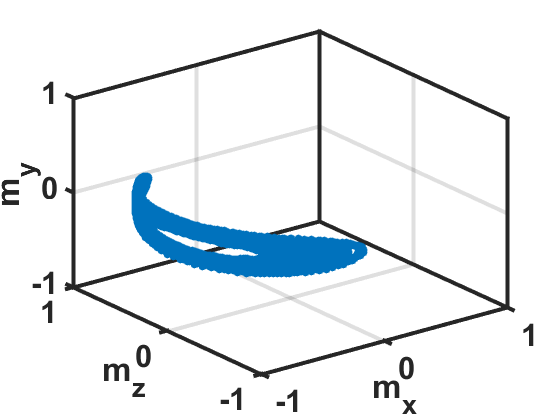}	
	}\subfigure[]{\includegraphics[width=2in]{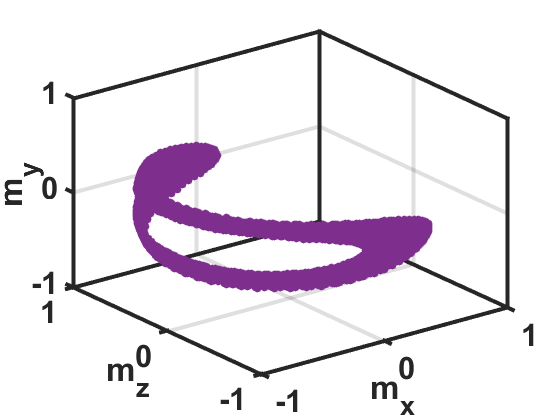}
	}\subfigure[]{\includegraphics[width=2in]{magnetization_V_dot14_normH_100.png}}
	\caption{Magnetization dynamics of trilayer-MTJ device under (a) an applied bias of $V=0.10V$, (b) $V=0.14V$ and (c) $V=0.16V$ and the applied magnetic field of $100$Oe}
	\label{magnetization_dynamics}
\end{figure*}
\indent We show in Fig.~\ref{magnetization_dynamics} the magnetization dynamics of the trilayer device under different applied biases. It can be seen from the Fig.~\ref{magnetization_dynamics} that as the bias voltage increases, the out of the plane component of the magnetization increases ($m_y$) due to the large spin current which results in a high frequency of oscillations. Also, it can be seen from the Fig.~\ref{magnetization_dynamics} that the spread in the magnetization dynamics due to thermal noise reduces with the bias voltage resulting in small line width of oscillations. 
\bibliographystyle{apsrev}
\bibliography{sample}
\end{document}